\documentclass[
reprint,
amsmath,amssymb,
aps,
pra,
]{revtex4-2}

\usepackage{graphicx, color}
\usepackage{dcolumn}
\usepackage{bm}
\usepackage{here}
\usepackage[whole]{bxcjkjatype}
\usepackage{enumitem}

\begin{document}

\title{Analyzing the optical pumping on the $5s4d\,{}^1D_2-5s8p\,{}^1P_1$ transition in a magneto-optical trap of Sr atoms}

\author{Naohiro Okamoto}
\author{Takatoshi Aoki}
\author{Yoshio Torii}
\email{ytorii@phys.c.u-tokyo.ac.jp}

\affiliation{Institute of Physics, The University of Tokyo, 3-8-1 Komaba, Meguro-ku, Tokyo 153-8902, Japan}
\date{\today}

\begin{abstract}

We explore the efficacy of optical pumping on the $5s4d\,{}^1D_2 - 5s8p\,{}^1P_1$ ($448\,\mathrm{nm}$) transition in a magneto-optical trap (MOT) of Sr atoms.  
The number of trapped atoms is enhanced by a factor of $12.0(6)$ relative to the case without repumping light, which is six times as large as that obtained using the pumping transition $5s4d\,{}^1D_2 - 5s6p\,{}^1P_1$ ($717\,\mathrm{nm}$).  
This enhancement is limited by decay pathways that bypass the $5s4d\,{}^1D_2$ state, namely $5s5p\,{}^1P_1 \to 5s4d\,{}^3D_1 \to 5s5p\,{}^3P_0$ and $5s5p\,{}^1P_1 \to 5s4d\,{}^3D_2 \to 5s5p\,{}^3P_2$, which account for 8\% of the total loss of the trapped atoms.  
We determine the decay rates for the $5s5p\,{}^1P_1 \to 5s4d\,{}^3D_1$ and $5s5p\,{}^1P_1 \to 5s4d\,{}^3D_2$ transitions to be $66(6)\,\mathrm{s^{-1}}$ and $2.4(2)\times10^2\,\mathrm{s^{-1}}$, respectively. 
Furthermore, we experimentally demonstrate for the first time that, when the trap beam diameter is small, escape of atoms in the $5s4d\,{}^1D_2$ state, which has a relatively long lifetime of $400\,\mathrm{\mu s}$, becomes a dominant loss mechanism, and that the $448\,\mathrm{nm}$ pumping light effectively suppresses this escape.  
Our findings will contribute to improved laser cooling and fluorescence imaging in cold strontium atom platforms, such as quantum computers based on optical tweezer arrays.

\end{abstract}

\maketitle

\section{introduction}

Alkaline-earth-metal(-like) atoms exhibit unique electronic structures characterized by long-lived metastable states and ultra-narrow optical transitions. 
These properties have made them a cornerstone in diverse areas of modern atomic physics, including precision metrology~\cite{S.L.Campbell2017,E.Oelker2019,T.L.Nicholson2015,W.F.McGrew2018,S.M.Brewer2019,T.Bothwell2019,N.Nemitz2016,BACONcolab2021, N.Dimarcq2024}, tests of special relativity~\cite{P.Delva2017}, gravitational redshift measurements~\cite{M.Takamoto2020,T.Bothwell2022,X.Zheng2023}, quantum simulation~\cite{S.Kolkowits2017}, quantum information~\cite{A.Daley2008, N.Schine2022, R.Tao2025}, gravitational wave detection~\cite{S.Kolkowits2016,M.Abe2021}, and search for dark matter~\cite{M.Abe2021,T.Kobayashi2022}.  

In a standard magneto-optical trap (MOT) of Sr atoms, laser cooling is initially performed on the $5s^2\,{}^1S_0 - 5s5p\,{}^1P_1$ transition at $461\,\mathrm{nm}$.  
This transition is not completely closed; a small fraction of atoms decays to the $5s4d\,{}^1D_2$ state, which subsequently decays to the long-lived $5s5p\,{}^3P_2$ state.
Conventionally, the $5s5p \,{}^3P_2 - 5s6s \,{}^3S_1$ ($707\,\mathrm{nm}$) transition is used to repump atoms in the $5s5p \,{}^3P_2$ state. 
Because atoms excited to the $5s6s \,{}^3S_1$ state can decay to the long-lived $5s5p \,{}^3P_0$ state{\,}\cite{A.V.Taichenachev2006, Z.W.Barber2006}, another laser at the $5s5p \,{}^3P_0 - 5s6s \,{}^3S_1$ ($679\,\mathrm{nm}$) transition{\,}\cite{K.R.Vogel1999, X.Xu2003} is necessary. 
Single-repumping schemes have also been demonstrated, specifically using $5s5p \,{}^3P_2 - 5s5d \,{}^3D_2$ ($497\,\mathrm{nm}$){\,}\cite{N.Poli2005}, $5s5p \,{}^3P_2 - 5s6d \,{}^3D_2$ ($403\,\mathrm{nm}$){\,}\cite{S.Stellmer2014}, $5s5p \,{}^3P_2 - 5p^2 \,{}^3P_2$ ($481\,\mathrm{nm}$){\,}\cite{F.Hu2019}, and $5s5p \,{}^3P_2 - 5s4d \,{}^3D_2$ ($3012 \,\mathrm{nm}$){\,}\cite{P.G.Mickelson2009}.

In recent years, experiments utilizing arrays of individually trapped Sr atoms in optical tweezers have gained significant attention~\cite{N.Schine2022, R.Tao2025, M.A.Norcia2019, A.Cooper2018, M.A.Norcia2018, A.Young2020}. 
In such experiments, fluorescence detection is typically performed using the $461\,\mathrm{nm}$ transition of Sr. 
However, once the atoms decay to the $5s4d\,{}^1D_2$ state, they remain in this dark state for an extended period ($\sim 400\,\mathrm{\mu s}$)~\cite{D.Husain1988}, which potentially reduces the fidelity of fluorescence detection.

To avoid this issue in fluorescence detection, a fast repumping scheme using the $5s4d\,{}^1D_2 - 5s8p\,{}^1P_1$ transition at $448\,\mathrm{nm}$ has recently been proposed~\cite{J.Samland2024}.  
In Ref.~\cite{J.Samland2024}, a $60\%$ increase in atomic flux of a two-dimensional MOT is reported.
However, the performance of repumping via the $448\,\mathrm{nm}$ transition in a three-dimensional (3D) MOT has not yet been investigated.

Another candidate for repumping the $5s4d\,{}^1D_2$ state is the $5s4d\,{}^1D_2 - 5s6p\,{}^1P_1$ transition at $717\,\mathrm{nm}$, which has been studied previously.
In that case, the enhancement factor of the atom number in a 3D MOT was reported to be approximately two~\cite{T.Kurosu1992, K.R.Vogel1999, Y.Bidel2002}.
Ref.~\cite{T.Kurosu1992} attributed this limited enhancement to decay channels from the $5s5p\,{}^1P_1$ state to the $5s4d\,{}^3D_{1,2}$ states.

Recently, we evaluated the performance of single-repumping schemes ($481\,\mathrm{nm}$ and $497\,\mathrm{nm}$) and identified that the decay pathway $5s5p\,{}^1P_1 \to 5s4d\,{}^3D_1 \to 5s5p\,{}^3P_0$ imposes an upper limit on the achievable enhancement factor~\cite{N.Okamoto2024}.  
However, the $5s5p\,{}^1P_1 \to 5s4d\,{}^3D_2$ transition has not yet been observed experimentally.

In this paper, we evaluate the repumping performance of the $5s4d\,{}^1D_2 - 5s8p\,{}^1P_1$ ($448\,\mathrm{nm}$) transition in a 3D MOT of $\mathrm{{}^{88} Sr}$.
We achieve an enhancement factor of $12.0(6)$ relative to the case without repumping light, which is six times as large as that obtained using the $5s4d\,{}^1D_2 - 5s6p\,{}^1P_1$ ($717\,\mathrm{nm}$) pumping transition.  
This enhancement is limited by the decay pathways $5s5p\,{}^1P_1 \to 5s4d\,{}^3D_{1,2}$, as suggested in Ref.~\cite{T.Kurosu1992}, while the contribution from the decay of the upper state $5s8p\,{}^1P_1$ to the $5s5p\,{}^3P_J$ states is found to be negligible.  
We determine the decay rates for the $5s5p\,{}^1P_1 \to 5s4d\,{}^3D_1$ and $5s5p\,{}^1P_1 \to 5s4d\,{}^3D_2$ transitions to be $66(6)\,\mathrm{s^{-1}}$ and $2.4(2)\times10^2\,\mathrm{s^{-1}}$, respectively. 
Furthermore, we experimentally identify for the first time that, when the trap beam size is small, escape of atoms in the $5s4d\,{}^1D_2$ state from the trapping region becomes a dominant loss mechanism, and that the $448\,\mathrm{nm}$ light effectively suppresses this escape.  
Our findings will contribute to laser cooling and imaging performance of cold strontium atom systems, such as quantum computers based on optical tweezer arrays of strontium atoms.

\section{experimental setup}
\begin{figure}
	\begin{center}
		\includegraphics[width=86mm]{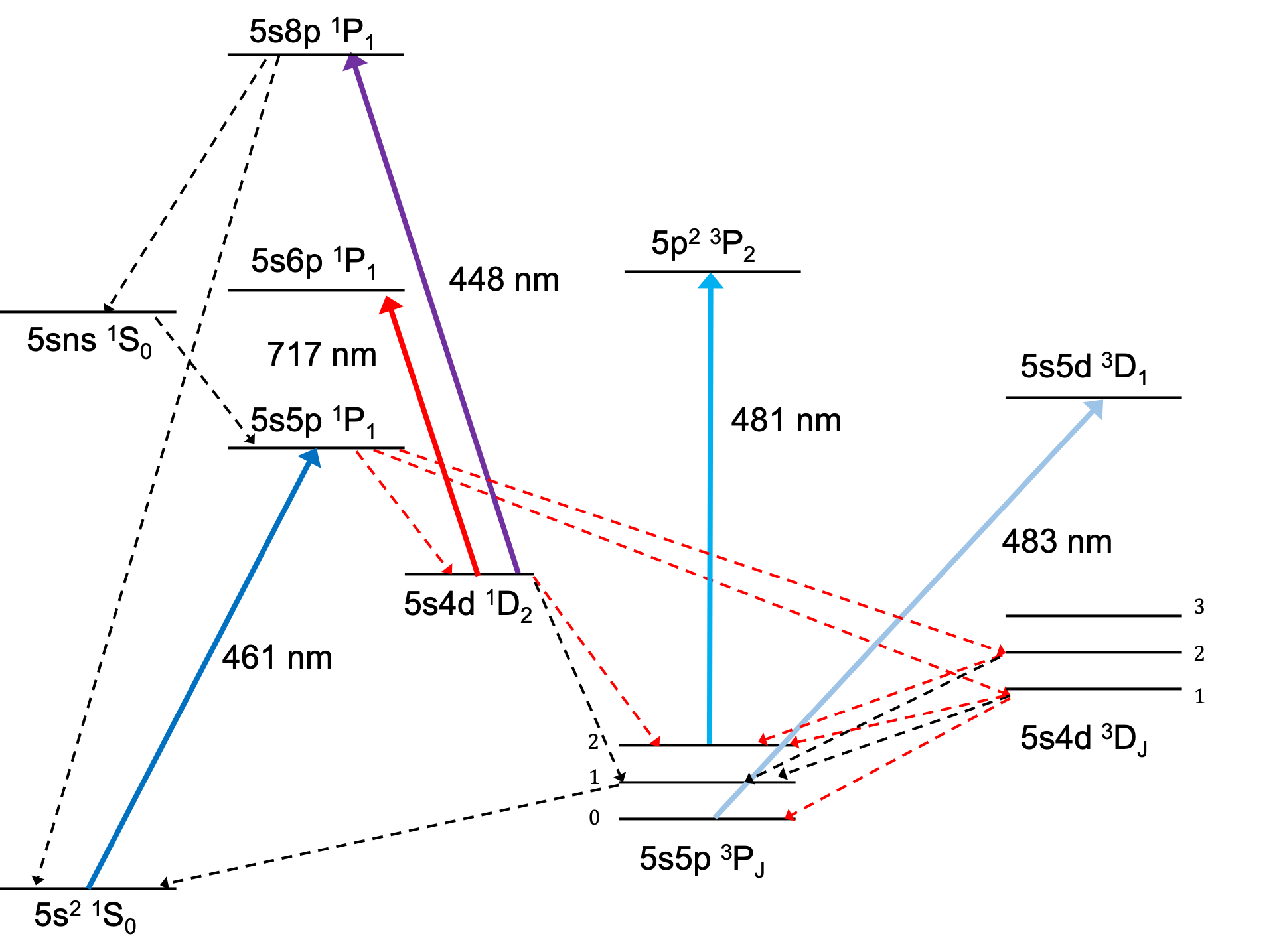}
		\caption{Energy level diagram of strontium relevant to this study.  Dashed arrows indicate decay processes, among which those highlighted in red represent decay pathways that lead to MOT loss.}
		\label{fig:energy_levels}
	\end{center}
\end{figure}

Figure~\ref{fig:energy_levels} shows the energy levels of strontium relevant to our investigation.  
The MOT operates on the $5s^2\,{}^1S_0 - 5s5p\,{}^1P_1$ transition at $461\,\mathrm{nm}$. 
The red dashed arrows in Fig.~\ref{fig:energy_levels} indicate four decay pathways that contribute to atom loss from the MOT as follows:
\begin{enumerate}[label=(\roman*)]
    \item $5s5p\,{}^1P_1\to5s4d\,{}^1D_2\to5s5p\,{}^3P_2$\label{exp:case1}
    \item $5s5p\,{}^1P_1\to5s4d\,{}^3D_1\to5s5p\,{}^3P_0$\label{exp:case2}
    \item $5s5p\,{}^1P_1\to5s4d\,{}^3D_2\to5s5p\,{}^3P_2$\label{exp:case3}
    \item $5s5p\,{}^1P_1\to5s4d\,{}^3D_1\to5s5p\,{}^3P_2$\label{exp:case4}
\end{enumerate}
To repump atoms in the $5s5p\,{}^3P_2$ state, we apply laser light resonant with the $5s5p\,{}^3P_2 - 5p^2\,{}^3P_2$ transition at $481\,\mathrm{nm}$.  
For atoms in the $5s5p\,{}^3P_0$ state, repumping is achieved using light resonant with the $5s5p\,{}^3P_0 - 5s5d\,{}^3D_1$ transition at $483\,\mathrm{nm}$.  
The $481\,\mathrm{nm}$ laser suppresses the loss due to the decay pathways \ref{exp:case1}, \ref{exp:case3}, and \ref{exp:case4}, while the $483\,\mathrm{nm}$ laser suppresses the loss due to the decay pathway \ref{exp:case2}.
To optically pump atoms in the $5s4d\,{}^1D_2$ state, we use the $5s4d\,{}^1D_2 - 5s8p\,{}^1P_1$ transition at $448\,\mathrm{nm}$.  
The atoms in the upper state of this transition decays to the ground state $5s^2\,{}^1S_0$ with a lifetime of approximately $50\,\mathrm{ns}$~\cite{H.G.C.Werij1992},  
whereas the atoms in the lower state, $5s4d\,{}^1D_2$, decays to the $5s5p\,{}^3P_J$ states with a lifetime of $400\,\mathrm{\mu s}$~\cite{D.Husain1988}.  
Therefore, when the $448\,\mathrm{nm}$ transition is saturated with sufficient intensity, the effective branching ratio to the $5s5p\,{}^3P_2$ state is on the order of $10^{-4}$, making the associated loss effectively negligible.  
The $448\,\mathrm{nm}$ laser thus serves to block the decay pathway \ref{exp:case1}.  
The roles of each pumping light are summarized in Table~\ref{table:blocking}.
\begin{table}[t]
    \caption{Decay paths suppressed by each repumping laser. A circle symbol indicates that the corresponding decay pathway is suppressed by the optical pumping light.}
    \label{table:blocking}
    \centering
    \begin{tabular}{c|ccc}
    \hline\hline
      & \multicolumn{3}{c}{Pumping transition} \\
    Decay path from $5s5p\,{}^1P_1$ & 481\,nm & 483\,nm & 448\,nm \\
    \hline
    $\to 5s4d\,{}^1D_2 \to 5s5p\,{}^3P_2$ & $\circ$ &  & $\circ$ \\
    $\to 5s4d\,{}^3D_1 \to 5s5p\,{}^3P_0$ &  & $\circ$ &  \\
    $\to 5s4d\,{}^3D_2 \to 5s5p\,{}^3P_2$ & $\circ$ &  &  \\
    $\to 5s4d\,{}^3D_1 \to 5s5p\,{}^3P_2$ & $\circ$ &  &  \\
    \hline\hline
    \end{tabular}
\end{table}
Applying various combinations of repumping lasers allows us to estimate the individual contributions of each decay pathway to the MOT loss rate.

The experimental setup is essentially the same as in our previous work~\cite{N.Okamoto2024}, except for the addition of a repumping beam at $448\,\mathrm{nm}$.  
The $448\,\mathrm{nm}$ beam has a power of $38\,\mathrm{mW}$ and a beam diameter of $2\,\mathrm{mm}$ ($I\sim10^3\,\mathrm{mW/cm^2}$), covering the entire MOT cloud.  
The saturation intensity for this transition is $I_s=\pi hc\gamma/(3\lambda^2)=4.4\,\mathrm{mW/cm^2}$, where $\gamma=2\pi\times3\,\mathrm{MHz}$~\cite{J.Samland2024}, $\lambda$ is wavelength, $c$ is the speed of light, and $h$ is the Planck constant.
Therefore, the beam is sufficiently intense to saturate the transition.  
All laser beams are generated from homemade external-cavity diode lasers.

The MOT consists of three retro-reflected beams at $461\,\mathrm{nm}$ with a beam diameter of $18\,\mathrm{mm}$ and a total power of $65\,\mathrm{mW}$.  
The axial magnetic field gradient is $50\,\mathrm{G/cm}$.  
The detuning of the trapping light is adjusted within the range of $-26\,\mathrm{MHz}$ to $-56\,\mathrm{MHz}$.  
The excitation fraction of the $5s5p\,{}^1P_1$ state at each detuning is determined using the method described in Appendix~A.  
The frequencies of all repumping lasers are set to resonance.  
For the lasers other than the $448\,\mathrm{nm}$ beam, frequency stabilization is achieved using a hollow cathode lamp~\cite{T.Sato2022}.
The frequency of the $448\,\mathrm{nm}$ laser is optimized by maximizing the MOT fluorescence while simultaneously introducing the $461\,\mathrm{nm}$ and $448\,\mathrm{nm}$ beams.

The atoms are loaded in the MOT directly from a thermal atomic beam derived from an oven with capillaries~\cite{M.Schioppo2012, N.Okamoto2025}.
The atoms are trapped in a glass cell ($25\,\mathrm{mm}\,\times\,25\,\mathrm{mm}\,\times\,100\,\mathrm{mm}$), and the entire vacuum system is evacuated by a single 55-l/s ion pump.
The oven temperature is set to $335\,\mathrm{{}^\circ C}$, resulting in a MOT loading rate of $5 \times 10^5\,\mathrm{atoms/s}$ and a vacuum pressure of $\sim 1 \times 10^{-10}\,\mathrm{Torr}$. 
At this vacuum pressure, the loss rate due to background gas collisions is $0.06\,\mathrm{s^{-1}}$~\cite{N.Okamoto2025}.

\section{results and discussion}

\begin{figure}
	\begin{center}
		\includegraphics[width=86mm]{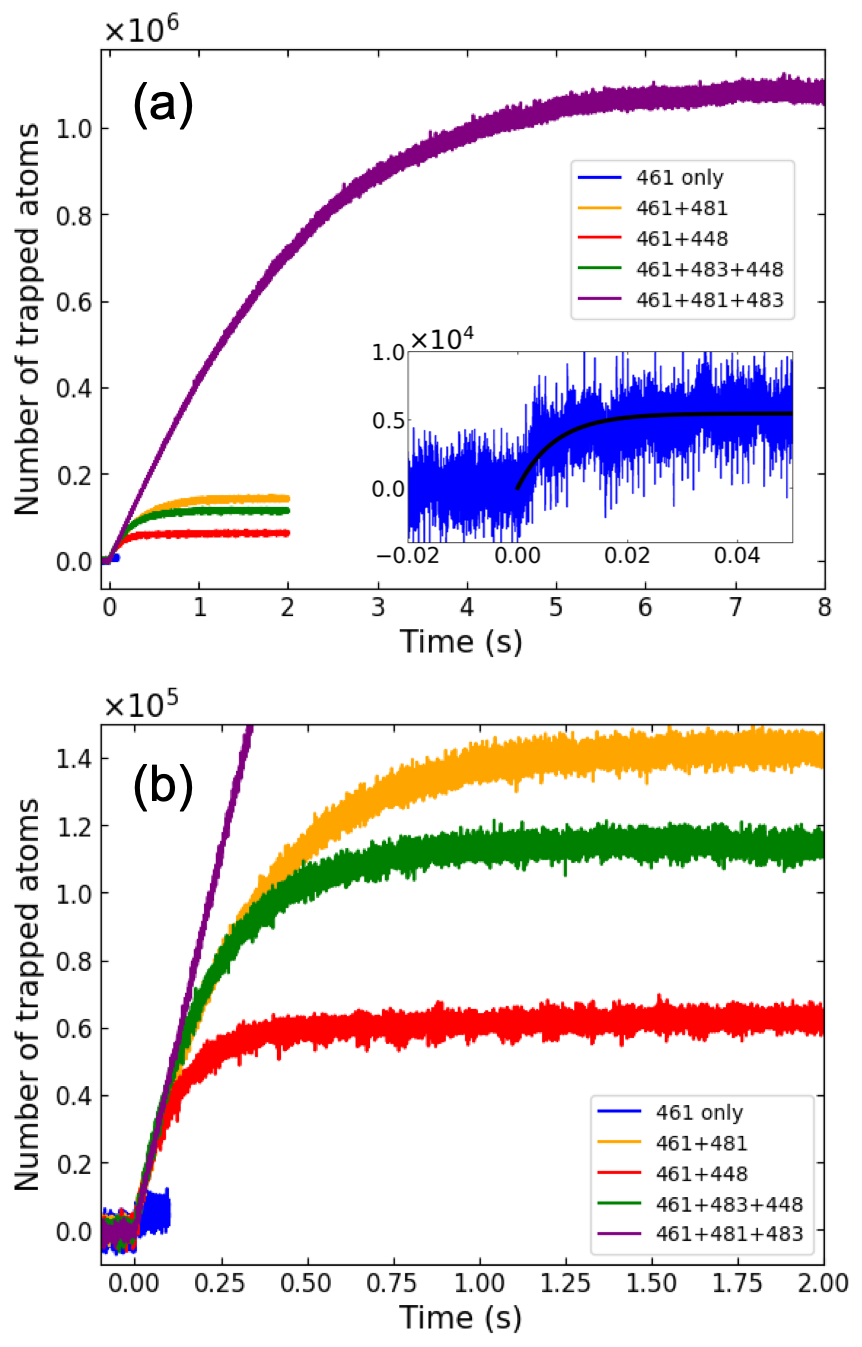}
		\caption{(a) Loading curves of the MOT under different repumping schemes. The detuning of the MOT beams is set to $-36\,\mathrm{MHz}$. The inset is an enlarged view of the loading curve for no repumping scheme (461\,nm only) with a fit to the data. (b) An enlarged view of all the loading curves.}
		\label{fig:loading}
	\end{center}
\end{figure}
As shown in Fig.~~\ref{fig:loading}, we measure the loading curves under the following repumping schemes: 
\begin{enumerate}
    \item 461\,nm (no repumping) \label{case0}
    \item 461\,nm + 481\,nm \label{case1}
    \item 461\,nm + 448\,nm \label{case2}
    \item 461\,nm + 483\,nm + 448\,nm \label{case4}
    \item 461\,nm + 481\,nm + 448\,nm \label{case5}
    \item 461\,nm + 481\,nm + 483\,nm (dual repumping) \label{case6}
\end{enumerate}
The loading curve for case \ref{case5} (461\,nm + 481\,nm + 448\,nm) is not included in Fig.~\ref{fig:loading} as it coincides exactly with that of case \ref{case1} (461\,nm + 481\,nm) and is therefore indistinguishable.

For case \ref{case6} (dual repumping scheme), all the decay paths from the $5s5p\,{}^1P_1$ state are suppressed and the number of trapped atoms is limited by two-body collisions~\cite{N.Okamoto2025}. 
On the other hand, for cases \ref{case0} - \ref{case5}, the atom densities are sufficiently low that the losses due to two-body collisions can be neglected, and the rate equation for the MOT atom number $N$ and the loss rate $L_X$ when applying laser light at wavelength(s) $X\,$\,(in nm) can be written as  
\begin{equation}
    \frac{dN}{dt} = R - L_X N, \label{eq:rate}
\end{equation}
where $R$ denotes the loading rate.  
In the steady state, the atom number $N_X$ is given by  
\begin{equation}
    N_X = \frac{R}{L_X}. \label{eq:N}
\end{equation}
This relation indicates that the steady-state atom number is inversely proportional to the loss rate.  
Based on this, we define the enhancement factor $\epsilon_X$ relative to the steady-state atom number obtained with only the $461\,\mathrm{nm}$ light as  
\begin{equation}
    \epsilon_X = \frac{N_X}{N_{461}} = \frac{L_{461}}{L_X}. \label{eq:enhancement}
\end{equation}
The enhancement factors for each repumping scheme, obtained from the data in Fig.~\ref{fig:loading}, are summarized in Table~\ref{table:enhancement}.
The uncertainties of the enhancement factors arise primarily from the uncertainty in the trapped atom number for case 1 (461\,nm only) [see the inset of Fig.~\ref{fig:loading} (a)].
\begin{table}[h]
    \caption{Enhancement factors of the atom number for each repumping scheme.}
    \label{table:enhancement}
    \centering
    \begin{tabular}{ccc}
    \hline\hline
    No. & laser wavelength(s) & enhancement factor \\ \hline
    1 & 461\,nm (no repumping) & $1$ \\
    2 & 461\,nm + 481\,nm & $26(1)$ \\
    3 & 461\,nm + 448\,nm & $12.0(6)$ \\
    4 & 461\,nm + 483\,nm + 448\,nm & $21(1)$ \\
    5 & 461\,nm + 481\,nm + 448\,nm & $26(1)$ \\
    6 & 461\,nm + 481\,nm + 483\,nm & $2.0(1)\times10^2$ \\
    \hline\hline
    \end{tabular}
\end{table}

\begin{figure}[h]
	\begin{center}
		\includegraphics[width=86mm]{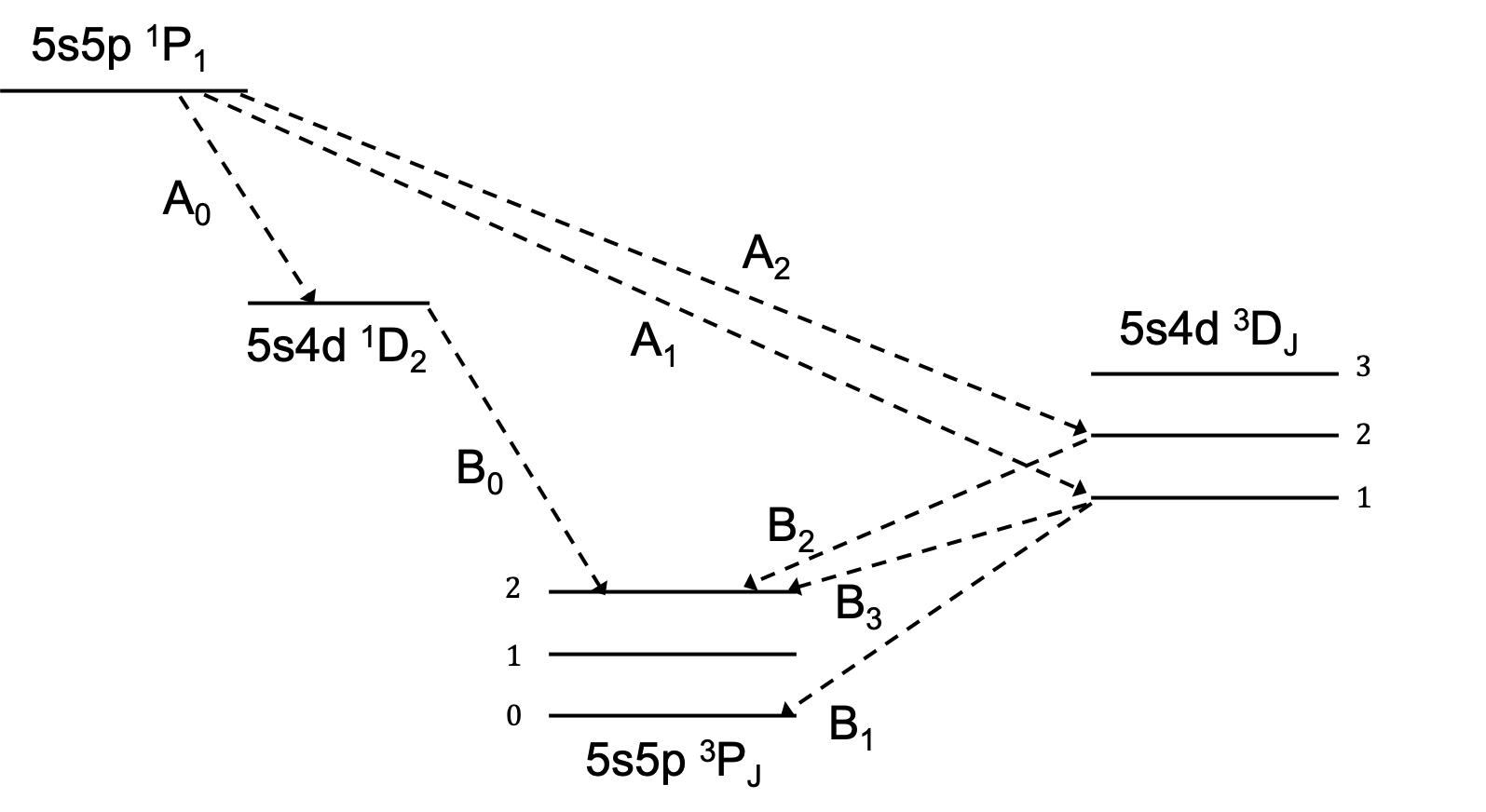}
		\caption{Decay rates and branching ratios relevant to MOT loss.}
		\label{fig:decay_path}
	\end{center}
\end{figure}

If we ignore decay from the upper state $5s8p\,{}^1P_1$ (via intermediate states) to the $5s5p\,{}^3P_{2,0}$ states, the loss rates for each repumping scheme are expressed as
\begin{align}
    L_{461} &= f(A_0B_0 + A_1B_1 + A_2B_2 + A_1B_3), \label{eq:461} \\
    L_{461+481} &= fA_1B_1, \label{eq:461+481} \\
    L_{461+448} &= f(A_1B_1 + A_2B_2 + A_1B_3), \label{eq:461+448} \\
    L_{461+483+448} &= f(A_2B_2 + A_1B_3), \label{eq:461+483+448} \\
    L_{461+481+448} &= fA_1B_1, \label{eq:461+481+448}
\end{align}
where $f$ denotes the excitation fraction of the $5s5p\,{}^1P_1$ state;  
$A_0$, $A_1$, and $A_2$ are the decay rates of $5s5p\,{}^1P_1 \to 5s4d\,{}^1D_2$, $5s5p\,{}^1P_1 \to 5s4d\,{}^3D_1$, and $5s5p\,{}^1P_1 \to 5s4d\,{}^3D_2$, respectively;  
and $B_0$, $B_1$, $B_2$, and $B_3$ are the branching ratios of $5s4d\,{}^1D_2 \to 5s5p\,{}^3P_2$, $5s4d\,{}^3D_1 \to 5s5p\,{}^3P_0$, $5s4d\,{}^3D_2 \to 5s5p\,{}^3P_2$, and $5s4d\,{}^3D_1 \to 5s5p\,{}^3P_2$, respectively (Fig.~\ref{fig:decay_path}).

From Eqs.~\eqref{eq:461+481}, \eqref{eq:461+448}, and \eqref{eq:461+483+448}, we obtain  
\begin{equation}
    L_{461+448} = L_{461+481} + L_{461+483+448}. \label{eq:relation1}
\end{equation}
This equation is expressed in terms of enhancement factor as
\begin{equation}
    \epsilon_{461+448} = \left[\epsilon_{461+481}^{-1} + \epsilon_{461+483+448}^{-1}\right]^{-1}. \label{eq:relation2}
\end{equation}
From the enhancement factors presented in Table~\ref{table:enhancement}, the left-hand side of Eq.~\eqref{eq:relation2} is 12.0(6), whereas the right-hand side of Eq.~\eqref{eq:relation2} is 11.7(4).
This shows that Eq.~\eqref{eq:relation2} holds within uncertainty, supporting the validity of our model that decay from the upper state $5s8p\,{}^1P_1$ to the $5s5p\,{}^3P_{2,0}$ states is negligible. 
This assumption is also supported by the fact that the enhancement factor for case \ref{case1} (461\,nm + 481\,nm) coincides with that for case \ref{case5} (461\,nm + 481\,nm + 448\,nm).
We suspect that the limited enhancement factor of about 2 for the $5s4d\,{}^1D_2 - 5s6p\,{}^1P_1$ ($717\,\mathrm{nm}$) repumping~\cite{T.Kurosu1992, K.R.Vogel1999, Y.Bidel2002} is due to losses from the upper state $5s6p\,{}^1P_1$ to the $5s5p\,{}^3P_{2,0}$ states.

According to Ref.~\cite{M.S.Safronova2013}, the branching ratios are calculated as $B_1=0.5953$, $B_2=0.1942$, and $B_3=0.0185$ from the theoretically evaluated line strengths.  
Using $B_3/B_1\sim0.03$, Eqs.~\eqref{eq:461}, \eqref{eq:461+448}, \eqref{eq:461+483+448}, and the experimentally obtained enhancement factors in Table~\ref{table:enhancement},  
we determine the relative contributions of each decay path, as summarized in Table~\ref{table:ratio}.
\begin{table}[t]
    \caption{Relative contributions of the decay paths derived from the enhancement factors in Table~\ref{table:enhancement} and the theoretically evaluated branching ratios~\cite{M.S.Safronova2013}.}
    \label{table:ratio}
    \centering
    \begin{tabular}{ccc}
    \hline\hline 
    Decay path from $5s5p\,{}^1P_1$ & Rate & Ratio \\ \hline
    $\to 5s4d\,{}^1D_2 \to 5s5p\,{}^3P_2$ & $A_0B_0$ & 91.4(3)\% \\
    $\to 5s4d\,{}^3D_1 \to 5s5p\,{}^3P_0$ & $A_1B_1$ & 3.8(1)\% \\
    $\to 5s4d\,{}^3D_2 \to 5s5p\,{}^3P_2$ & $A_2B_2$ & 4.6(2)\% \\
    $\to 5s4d\,{}^3D_1 \to 5s5p\,{}^3P_2$ & $A_1B_3$ & 0.120(5)\% \\
    \hline\hline
    \end{tabular}
\end{table}
The decay pathways that bypass the $5s4d\,{}^1D_2$ state, namely $5s5p\,{}^1P_1 \to 5s4d\,{}^3D_1 \to 5s5p\,{}^3P_0$ and $5s5p\,{}^1P_1 \to 5s4d\,{}^3D_2 \to 5s5p\,{}^3P_2$, account for $\sim 8\%$ of the total loss from the MOT, which limits the enhancement factor to $\sim 12$ for the $448\,\mathrm{nm}$ repumping.

\begin{figure}
	\begin{center}
		\includegraphics[width=86mm]{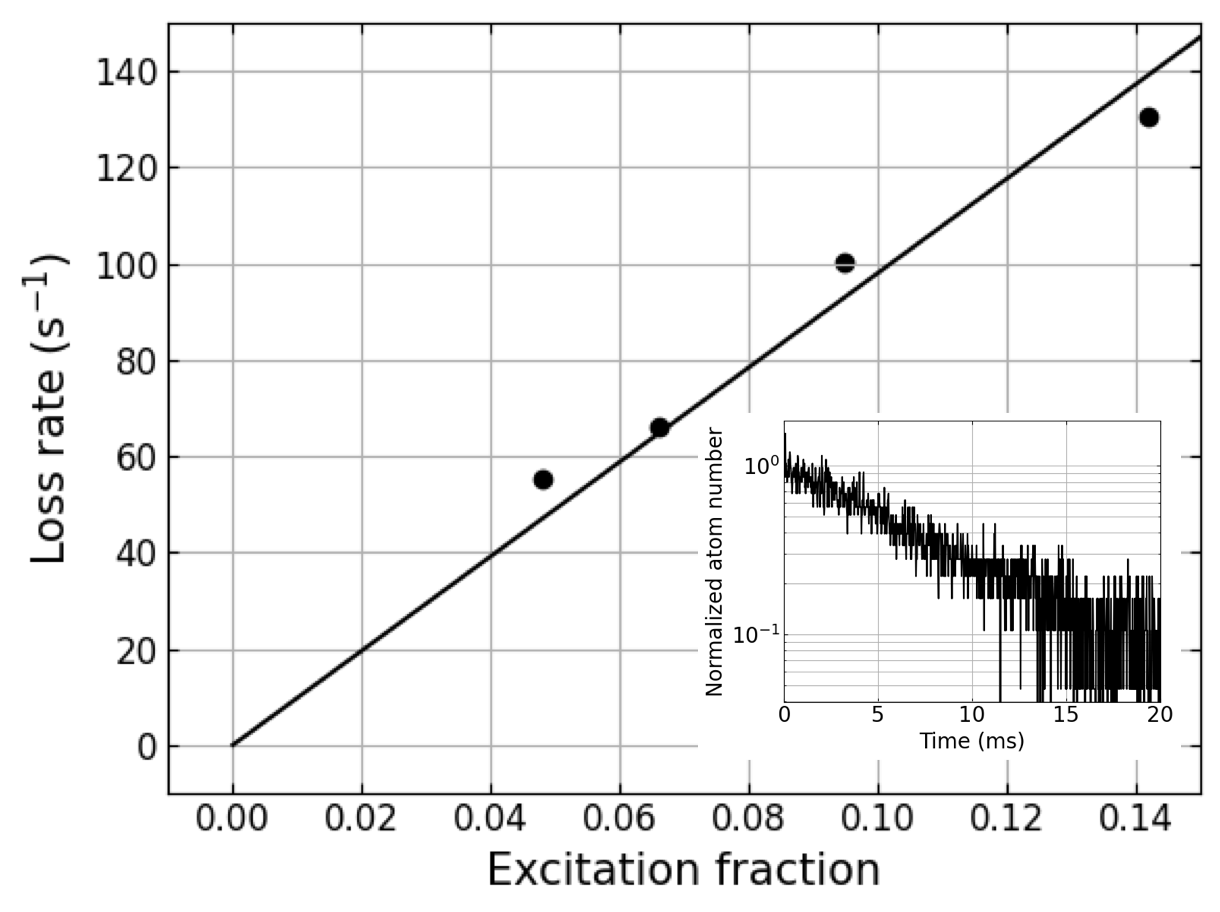}
		\caption{Dependence of the MOT loss rate on the excitation fraction of the $5s5p\,{}^1P_1$ state.  The inset shows the MOT decay at a detuning of $-26\,\mathrm{MHz}$ ($f=0.14$).}
		\label{fig:f_tau}
	\end{center}
\end{figure}

The transition rates $A_1$ and $A_2$ can be determined from the dependence of the MOT loss rate on the excitation fraction $f$ of the $5s5p\,{}^1P_1$ state. 
Figure~\ref{fig:f_tau} shows the dependence of the MOT loss rate on $f$, when the $481\,\mathrm{nm}$ laser is turned off in the case of dual repumping scheme ($461\,\mathrm{nm}+481\,\mathrm{nm}+483\,\mathrm{nm}$).
The MOT loss rate is then expressed as  
\begin{equation}
    L_{461+483} = f(A_0B_0 + A_2B_2 + A_1B_3). \label{eq:MOT_loss}
\end{equation}
Thus, from the slope in Fig.~\ref{fig:f_tau}, we obtain
\begin{equation}
    A_0B_0 + A_2B_2 + A_1B_3 = 9.8(4)\times10^2\,\mathrm{s^{-1}}, \label{eq:MOT_tau}
\end{equation}
where the uncertainty comes from the least-squares fit.
Using the branching ratios $B_1$ and $B_2$ in Ref.~\cite{M.S.Safronova2013}, together with the results in Table~\ref{table:ratio} and Eq.~\eqref{eq:MOT_tau}, we determine $A_1=66(6)\,\mathrm{s^{-1}}$ and $A_2=2.4(2)\times10^2\,\mathrm{s^{-1}}$.  
The value of $A_1$ is consistent with our previous result, $A_1=83(32)\,\mathrm{s^{-1}}$~\cite{N.Okamoto2024}.  
In the previous work, the uncertainty in $A_1$ was largely determined by the relatively large uncertainty in the literature value of $A_0$~\cite{L.R.Hunter1986}.  
In the present work, the uncertainty in $A_1$ has been significantly reduced because it is derived without relying on the literature value of $A_0$.

We also determine $A_0B_0=9.3(9)\times10^2\,\mathrm{s^{-1}}$. 
Using the commonly quoted value $B_0=0.32$ from Ref.~\cite{C.W.Bauschlicher1985}, this yields $A_0=2.8(3)\times10^3\,\mathrm{s^{-1}}$.  
This value is consistent with the experimental value reported in Ref.~\cite{L.R.Hunter1986}, $3.9(1.5)\times10^3\,\mathrm{s^{-1}}$,  
but significantly deviates from the theoretical value in Ref.~\cite{A.Cooper2018}, $9.25(40)\times10^3\,\mathrm{s^{-1}}$.  
A possible reason for this discrepancy is that the theoretical calculation of $B_0$ in Ref.~\cite{C.W.Bauschlicher1985} is significantly inaccurate.  
We intend to address this issue in future work.

\begin{figure}
	\begin{center}
		\includegraphics[width=86mm]{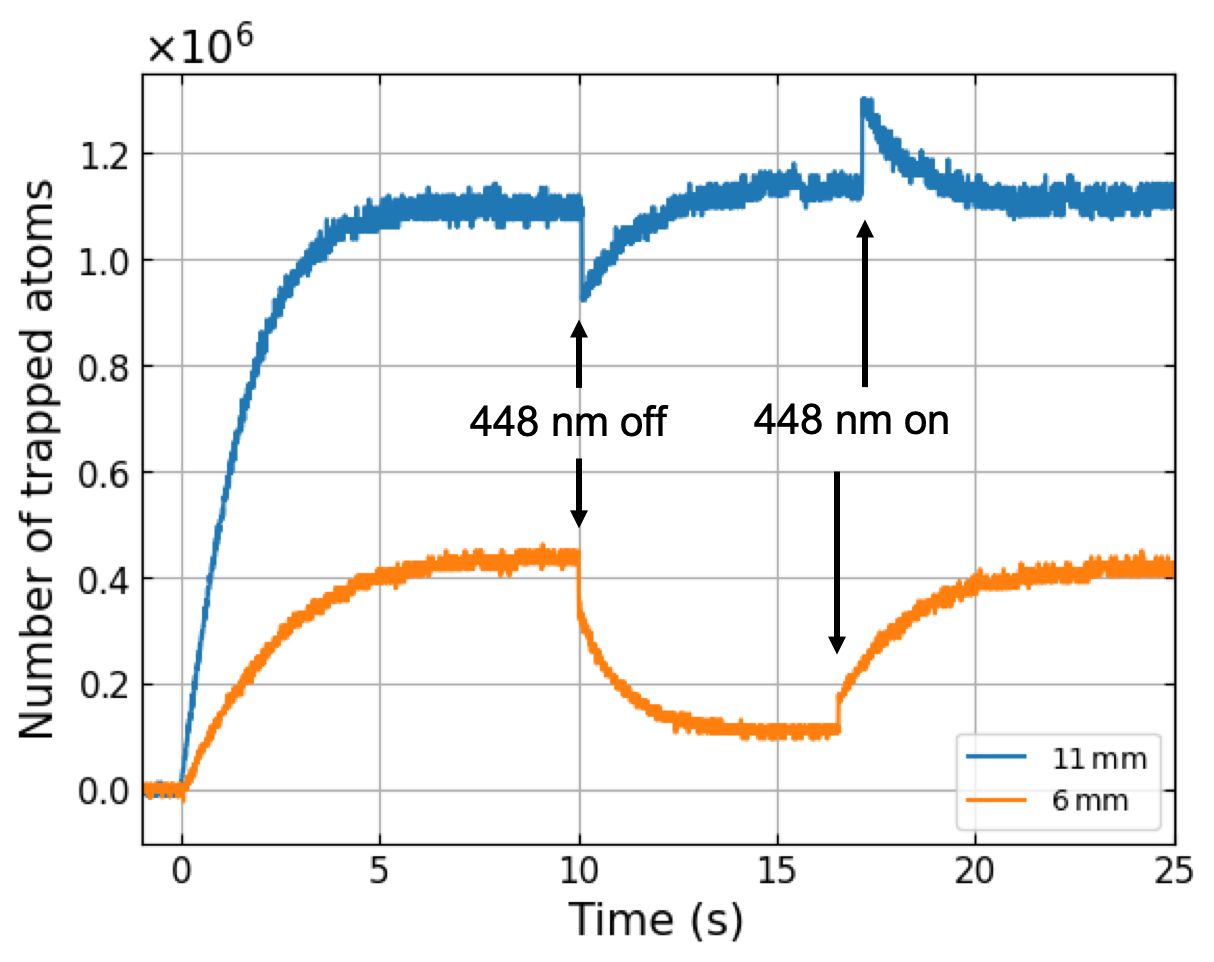}
		\caption{Evolution of the trapped atom number for MOT beam diameters of $11\,\mathrm{mm}$ and $6\,\mathrm{mm}$.  Initially, the MOT is loaded with the ${}^3P_2$ repumping light at $481\,\mathrm{nm}$, the ${}^3P_0$ repumping light at $483\,\mathrm{nm}$, and the ${}^1D_2$ optical pumping light at $448\,\mathrm{nm}$. At $t=10\,\mathrm{s}$, the $448\,\mathrm{nm}$ light is switched off. After the number of atoms stabilizes, the $448\,\mathrm{nm}$ light is turned on again. The detuning of the MOT beams is set to $-36\,\mathrm{MHz}$.}
		\label{fig:1D2_escape}
	\end{center}
\end{figure}
As a demonstration of the effectiveness of fast repumping on the $5s4d\,{}^1D_2$ state, we investigate the regime where escape of atoms in the dark $5s4d\,{}^1D_2$ state from the trapping region becomes a dominant loss mechanism when the trap beam diameter is reduced.  
Although such escape from the trapping region has been mentioned in some literatures~\cite{K.R.Vogel1999, C.Vishwakarma2019, T.Kurosu1992}, it has not been directly verified experimentally.  
Figure~\ref{fig:1D2_escape} shows the loading curves of the MOT atom number for trap beam diameters of $11\,\mathrm{mm}$ and $6\,\mathrm{mm}$.  
In these measurements, the MOT is initially loaded with the ${}^3P_2$ repumping light at $481\,\mathrm{nm}$, the ${}^3P_0$ repumping light at $483\,\mathrm{nm}$, and the ${}^1D_2$ repumping light at $448\,\mathrm{nm}$.  
Subsequently, the $448\,\mathrm{nm}$ light is turned off and then reintroduced.

In the case of the trap beam diameter of $11\,\mathrm{mm}$, when the $448\,\mathrm{nm}$ light is switched off, atoms accumulate in the $5s4d\,{}^1D_2$ state, causing the number of trapped atoms $N$ to decrease sharply. 
Afterwards, $N$ gradually recovers to its original level because two-body collisional losses are reduced
(the steady-state atom number without 448\,nm pumping light is actually slightly increased because of the slightly increased volume of the atom cloud due to time of flight of atoms in the $5s4d\,{}^1D_2$ state).
When the $448\,\mathrm{nm}$ light is turned back on, atoms in the $5s4d\,{}^1D_2$ state are quickly pumped back into the cooling cycle, resulting in a rapid increase in $N$.  
This increase enhances two-body collisions, causing $N$ to decrease again to its original level.

In the case of the trap beam diameter of $6\,\mathrm{mm}$, the behavior is quite different.
When the $448\,\mathrm{nm}$ light is turned off, as in the case of the trap beam diameter of $11\,\mathrm{mm}$, atoms decay into the $5s4d\,{}^1D_2$ state, leading to a sharp decrease in $N$. 
However, in contrast to the case with a beam diameter of $11\,\mathrm{mm}$, the further decrease in $N$ with a time constant of approximately $1\,\mathrm{s}$ is observed.
This decrease is attributed to the escape of atoms in the $5s4d\,{}^1D_2$ state.
Based on the decay rate of the $5s5p\,{}^1P_1 \to 5s4d\,{}^1D_2$ transition ($fA_0 \sim 400\,\mathrm{s^{-1}}$), the escape probability per ${}^1D_2$ atom is estimated to be 0.3\%, which is consistent with a numerical calculation assuming a Maxwell-Boltzmann distribution with the measured MOT temperature of $3\,\mathrm{mK}$~\cite{T.Kurosu1992}.
When the $448\,\mathrm{nm}$ light is reintroduced, atoms in the $5s4d\,{}^1D_2$ state are rapidly pumped back into the cooling cycle, causing $N$ to increase sharply.
Afterwards, as the escape of atoms in the $5s4d\,{}^1D_2$ state is suppressed, $N$ returns to its original level.

\section{conclusion}
In this work, we evaluated the repumping performance of the $5s4d\,{}^1D_2 - 5s8p\,{}^1P_1$ ($448\,\mathrm{nm}$) transition in a three-dimensional MOT.  
The enhancement in atom number was $12.0(6)$, limited by decay paths that bypass the $5s4d\,{}^1D_2$ state. 
We found that, in contrast to the $5s4d\,{}^1D_2 - 5s6p\,{}^1P_1$ ($717\,\mathrm{nm}$) repumping scheme, decay from the upper state $5s8p\,{}^1P_1$ to the $5s5p\,{}^3P_J$ states is negligible.  
The transition rates of $5s5p\,{}^1P_1 \to 5s4d\,{}^3D_1$ and $5s5p\,{}^1P_1 \to 5s4d\,{}^3D_2$ were determined to be $66(6)\,\mathrm{s^{-1}}$ and $2.4(2)\times10^2\,\mathrm{s^{-1}}$, respectively.  
Furthermore, we experimentally demonstrated for the first time that escape of atoms in the $5s4d\,{}^1D_2$ state becomes a dominant loss mechanism when the trap beam size is small, and that the $448\,\mathrm{nm}$ light effectively suppresses this escape. 
For large beam diameters, on the other hand, adding the 448\,nm light to the dual repumping scheme (481\,nm + 483\,nm or 707\,nm + 679\,nm) makes a negligible contribution to the increase in atom number.

At present, there is a significant discrepancy among literature values for the $5s5p\,{}^1P_1\to5s4d\,{}^1D_2$ transition rate $A_0$, which remains unsettled~\cite{J.Samland2024}.  
In future work, we aim to determine this rate more precisely by accurately measuring the branching ratio $B_0$ of the $5s4d\,{}^1D_2\to5s5p\,{}^3P_2$ transition.  
We plan to perform measurements of the branching ratio $B_0$ by analyzing the transient response of the MOT atom number under the application of $448\,\mathrm{nm}$ light, which will be published elsewhere.

\section{acknowledgments}
This work was supported by JSPS KAKENHI Grant Numbers 23K20849 and 22KJ1163.

\appendix

\section{Accurate Determination of the Excitation Fraction of the $5s5p\,{}^1P_1$ state}

The excitation fraction of the $5s5p\,{}^1P_1$ state in the $461\,\mathrm{nm}$ cooling transition is given by
\begin{equation}
    f\left( \delta \right) = \frac{1}{2} \frac{s_0}{1+s_0+4\left(\delta/\Gamma\right)^2}, \label{eq:append_f}
\end{equation}
where $\Gamma=2\pi\times30\,\mathrm{MHz}$ is the natural linewidth and $s_0 = I/I_s$ is the resonant saturation parameter. Here, $I$ denotes the total intensity of the $461\,\mathrm{nm}$ MOT beams and $I_s = 40\,\mathrm{mW/cm^2}$ is the saturation intensity.  
To accurately determine $f$, it is essential to evaluate $s_0$ precisely.  
The straightforward approach to determining $s_0$ is accurately measuring the total beam intensity $I$.
However, this method is challenging because of the uncertainties in gauging the shape of the beam and the loss of beam power in various optical components.

\begin{figure}[h]
	\begin{center}
		\includegraphics[width=86mm]{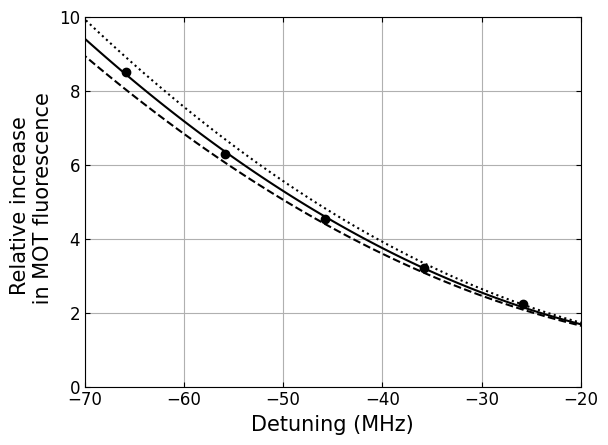}
		\caption{Relative increase in MOT fluorescence as a function of the initial trapping laser detuning. The dotted, solid , and dashed lines indicate theoretical curves based on Eq.\eqref{eq:append_power_ratio} with $s_0=1.4$, $1.55$, and $1.7$, respectively.}
		\label{fig:fitting}
	\end{center}
\end{figure}

To precisely determine $s_0$, we performed an experiment in which the detuning of the MOT beams was instantaneously set to zero for a short duration ($\sim 50\,\mathrm{\mu s}$) by ramping the injection current into the laser diode, and the relative increase in fluorescence was observed for various initial detunings (Fig.~\ref{fig:fitting}).
We confirmed that atom loss from the MOT is negligible within this timescale.

The steady-state fluorescence power of the MOT at a trapping laser detuning $\delta$, denoted as $P\left(\delta\right)$, can be expressed as  
\begin{equation}
    P\left(\delta\right) = \hbar \omega \Gamma N_0 f\left(\delta\right), \label{eq:append_MOT_fluorescence}
\end{equation}
where $N_0$ represents the steady-state atom number in the MOT and $\hbar \omega$ represents the photon energy.
From Eqs.~\eqref{eq:append_f} and \eqref{eq:append_MOT_fluorescence}, the relative increase in fluorescence is then given by  
\begin{equation}
    \frac{P\left(0\right)}{P\left(\delta\right)} = 1+\frac{4}{1+s_0}\left(\frac{\delta}{\Gamma}\right)^2. \label{eq:append_power_ratio}
\end{equation}
By fitting the data shown in Fig.~\ref{fig:fitting} to Eq.~\eqref{eq:append_power_ratio}, the resonant saturation parameter was determined to be $s_0=1.55(9)$.  
Using this value and Eq.~\eqref{eq:append_f}, the excitation fraction $f$ was accurately determined.

\bibliography{references.bib}

\end{document}